\documentclass[12pt]{article}

\usepackage{amsmath, amssymb, mathtools}
\usepackage{type1cm}
\usepackage{mathrsfs}
\usepackage{physics}
\usepackage[T1]{fontenc}                                                                                                                                                                                                                                                                                                                                                                                                                                                                                                                       
\usepackage{fancyhdr}
\usepackage{appendix}
\usepackage{titlesec}
\usepackage{cite}
\usepackage{color}
\usepackage[top=25truemm,bottom=25truemm,left=20truemm,right=20truemm]{geometry}
\usepackage[pdftex]{graphicx}
\usepackage{breqn}
\usepackage{comment}
\usepackage{braket}
\baselineskip=\normalbaselineskip

\makeatletter
 
 \@addtoreset{equation}{section}
\makeatother

\begin{document}
\setcounter{footnote}{0}
\setcounter{tocdepth}{3}
\bigskip
\def\thefootnote{\arabic{footnote}}

\begin{titlepage}
\renewcommand{\thefootnote}{\fnsymbol{footnote}}
\begin{normalsize}
\begin{flushright}
\begin{tabular}{l}
KOBE-COSMO-20-15\\
\end{tabular}
\end{flushright}
  \end{normalsize}

~~\\

\vspace*{0cm}
    \begin{Large}
       \begin{center}
         {$O(d,d;\mathbb{Z})$ invariant Fierz-Pauli massive gravity}
       \end{center}
    \end{Large}

\vspace{0.7cm}

\begin{center}
Toshifumi N\textsc{oumi}\footnote[1]
            {
e-mail address : tnoumi@phys.sci.kobe-u.ac.jp},
Kaishu S\textsc{aito}\footnote[2]
            {
e-mail address : 
184s151s@stu.kobe-u.ac.jp
},
Jiro S\textsc{oda}\footnote[3]
            {
e-mail address : 
jiro@phys.sci.kobe-u.ac.jp}
 and
Daisuke Y\textsc{oshida}\footnote[4]
            {
e-mail address : 
dyoshida@hawk.kobe-u.ac.jp
}

\vspace{0.7cm}

      {\it Department of Physics, Kobe University, Kobe 657-8501, Japan }
                                 
                      \end{center}

\vspace{0.5cm}

\begin{abstract}
\noindent
We consider an $O(d,d;\mathbb{Z})$ invariant massive deformation of double field theory at the level of free theory.
We study Kaluza-Klein reduction on $R^{1,n-1} \times T^{d}$ and derive the diagonalized second order action for each helicity mode. Imposing the absence
of ghosts and tachyons, we obtain a class of consistency conditions which include the well known weak constraint in double field theory as a special case.
Consequently, we find two-parameter sets of $O(d,d;\mathbb{Z})$ invariant Fierz-Pauli massive gravity theories.
\end{abstract}

\end{titlepage}

\section{Introduction}

Duality plays a central role in string theory, the most successful theory of quantum gravity. While it has boosted various  theoretical developments, its phenomenological implications have also been studied intensively. Especially, T-duality has interesting impacts on cosmology. For example, T-duality can be used to constrain higher derivative corrections~\cite{Veneziano:1991ek, Meissner:1991zj, Maharana:1992my, Bergshoeff:1995cg, Meissner:1996sa, Godazgar:2013bja, Tdual constraint, Eloy:2020dko}, based on which cosmological solutions have been studied incorporating all orders of $\alpha'$ corrections~\cite{HZ cosmo}. It may shed some light on a stingy realization of accelerated expansion of the universe beyond supergravity approximation~\cite{HZ cosmo,Bernardo1,Bernardo2}. Also, an interesting possibility has been explored that winding modes of the string may resolve cosmological singularities essentially as a consequence of T-duality~\cite{Kripfganz:1987rh, BVmechanism,Brandenberger:2008nx}.

Double Field Theory (DFT) is a field theoretic framework incorporating both the winding modes and the Kaluza-Klein (KK) modes of the string in a T-duality manifest fashion~\cite{DFT,Zwiebach Review,Gmetric formulation,Siegel:1993}, which would be useful, e.g., for exploring the aforementioned cosmological scenarios~\cite{WuYang1,WuYang2, Angus:2018mep, BrandenbergerDFT, Bernardo:2019pnq, Angus:2019bqs}. If we consider a theory with an internal $d$-dimensional torus, T-duality is captured by an $O (d,d;\mathbb{Z} )$ symmetry group which mixes winding modes and KK modes. The transformation rules of the background fields such as the metric and anti-symmetric $b$-field also follow in the standard manner. DFT is constructed to respect the $O (d,d;\mathbb{Z} )$ symmetry as well as appropriate gauge symmetries, which include the diffeomorphism symmetries of graviton and the gauge symmetry of $b$-field~\cite{DFT,Zwiebach Review,Gmetric formulation}.

In formulating DFT, it is nontrivial to maintain the gauge invariance. In the free theory, gauge invariance is guaranteed by imposing the weak constraint corresponding to the level-matching condition of the worldsheet theory. On the other hand, once we turn on interactions, the gauge invariance and the closure of gauge transformation are not guaranteed by the weak constraint alone. In the present formulation~\cite{Zwiebach Review,TypeII DFT,  Gmetric formulation, bg indi DFT}, the so-called strong constraint is imposed on top of the weak constraint in order to overcome these difficulties. However, the cost is that the winding modes are projected out by the strong constraint, so that one of the important stringy features is lost~\cite{bg indi DFT}: Ideally, we would like to have a consistent interacting DFT which accommodates both winding modes and KK modes by relaxing the strong constraint. 

Notably, in Ref.~\cite{The space time of DFT}, Holm and Zwiebach succeeded in making  the strong constraint in type II DFT~\cite{TypeII DFT} mild and partially incorporated winding modes of the R-R fields without spoiling the gauge invariance. They also showed that under the mild version of the strong constraint, type II DFT reduces to massive type IIA theory \cite{massive IIA}. Note that the NS-NS two-form is massive in massive type IIA theory, so that gauge invariance associated to the two-form is spontaneously broken. This motivates us to explore massive deformations of DFT as a bypass to phenomenology of winding modes: Since massive theories do not have gauge invariance from the beginning, it might be technically possible to formulate a consistent interacting theory without imposing the strong constraint.

In this paper, as a first step toward such a direction, we study massive deformations of DFT within the free theory. In particular, we find that a certain condition analogous to the standard weak constraint in the massless DFT is required for the theory to be free from ghosts and tachyons\footnote{
Of course, the strong constraint does not play any role in the present paper since we are focusing on the free theory as a first step toward massive deformations of the full interacting DFT. }. Note that such massive deformations will also be useful for exploring stringy UV completion of massive gravity, a phenomenological model of the accelerating expansion of the present universe~\cite{FierzPauli, deRham:2010ik, Resummation of MG}( see Refs.~\cite{Hinterbichler:2011tt, deRham:2014zqa} for reviews).

In the rest of the paper, we first review massless DFT on $R^{1,n-1} \times T^d$ (Sec.~2). Then, in Sec.~3, we study its massive deformations. There we consider a family of theories without imposing gauge invariance and the weak constraint corresponding to the level-matching condition, and then discuss consistency of the spectrum from the $n$-dimensional field theory point of view. We show that ghost and tachyon free conditions require a certain condition analogous to the weak constraint. Also we demonstrate that the standard weak constraint is picked up if we require in addition that the lightest massive spin $2$ particle is lighter than the string scale.
Note that there is a work by Olaf et al. on massive DFT, which showed that the level-matching condition is sufficient for the theory to be tachyon free at the level of equation of motion~\cite{Curious}. On the other hand, our paper derives necessary conditions for the theory to be free from ghosts and tachyons at the level of Lagrangian. 

\section{Review of Massless DFT}
In this section, we review the massless DFT on an $n$ dimensional Minkowski space with a $d$ dimensional torus, $R^{1,n-1} \times T^{d}$, following~\cite{DFT,Zwiebach Review}.
We choose the background metric $g^{bg}_{ij}$ as
\begin{align}
 g^{bg}_{ij}dx^{i} dx^{j} = \eta_{\mu\nu}dx^{\mu}dx^{\nu} + \delta_{ab}dx^{a} dx^{b},\qquad x^{a} \sim x^{a} + 2 \pi R^{a},
\end{align}
where we adopted the following index notations: $i,j,k,...$  for coordinates on the entire $D := n + d$ dimensional target spacetime $R^{1, n-1}\times T^{d}$, $\mu, \nu,\rho...$ for $R^{1,n-1}$, and $a,b,c,...$ for $T^{d}$. Here $\eta_{\mu\nu} = \text{diag}(-1,1,1, \dots ,1)$ is the Minkowski metric, $\delta_{ab}$ is the Kronecker delta, and $R^{a}$ is the radius of the $x^a$-cycle of the torus $T^d$.

DFT is a field theory which describes a gravitational field $h_{ij}$, an anti-symmetric $b$-field $b_{ij}$, and a dilaton field $d$ in a T-duality manifest way. Based on string theory on $R^{1,n-1} \times T^{d}$, each field is labeled by momentum $p_{a}$ and winding numbers $\tilde{p}^{a}$, which are quantized as
\begin{align}
 p_{a} =  \frac{1}{R^{a}} n_{a}, \qquad \tilde{p}^{a} = \frac{R^a}{\alpha'} w^{a} \qquad \left( n_{a}, w^{a} \in \mathbb{Z} \right)\,.
\end{align}
Here $\alpha'$ is the Regge slope, which is related with the string length $l_s$ through $\alpha' = (1/2) {l_{s}}^2$.
The key property of the string spectrum is that the momentum and the winding numbers obey the level-matching condition, which is also called the weak constraint in the context of DFT:
\begin{align}
 p_{a} \tilde{p}^{a} = 0.
\end{align}
The fields are labeled as $h_{ij}(x^{\mu}, p_{a}, \tilde{p}^{a}), b_{ij}(x^{\mu}, p_{a}, \tilde{p}^{a})$ and $d(x^{\mu}, p_{a}, \tilde{p}^{a})$. 

By construction, $p_{a}$ are the Fourier momentum dual to the coordinates $x^{a}$.
Similarly we introduce the dual coordinates $\tilde{x}_{a}$ of $\tilde{p}^{a}$ through the Fourier transformations. Thus we can describe $h_{ij}, b_{ij}$ and $d$ as fields living on the doubled space $\{x^{\mu}, x^{a}, \tilde{x}_{a}\}  \in R^{1,n-1} \times T^{d} \times T^{d} $: 
\begin{align}
 h_{ij}(x^{\mu}, x^{a}, \tilde{x}_{a}) &:= \sum_{n_{a}, w^{a} \in \mathbb{Z}} h_{ij}(x^{\mu}, p_{a}, \tilde{p}^{a}) \mathrm{e}^{i (p_{a} x^{a} + \tilde{p}^{a} \tilde{x}_{a})}, \notag\\
 b_{ij}(x^{\mu}, x^{a}, \tilde{x}_{a}) &:= \sum_{n_{a}, w^{a} \in \mathbb{Z}} b_{ij}(x^{\mu}, p_{a}, \tilde{p}^{a}) \mathrm{e}^{i (p_{a} x^{a} + \tilde{p}^{a} \tilde{x}_{a})}, \notag\\
 d(x^{\mu}, x^{a}, \tilde{x}_{a}) &:= \sum_{n_{a}, w^{a} \in \mathbb{Z}} d(x^{\mu}, p_{a}, \tilde{p}^{a}) \mathrm{e}^{i (p_{a} x^{a} + \tilde{p}^{a} \tilde{x}_{a})}. 
 \label{DoubleFourier}
\end{align}
Note that the dual coordinates enjoy the periodicity,
\begin{equation}
\tilde{x}_{a} \sim \tilde{x}_{a} + 2 \pi \frac{\alpha'}{R^{a}}.
 \label{identification of coordinate}
\end{equation}
Now the level-matching condition can be phrased as
\begin{equation}
\partial_{a} \tilde{ \partial}^{a} ( h_{ij}, b_{ij}, \text{and } d ) = 0.
\label{level matching}
\end{equation}
We also introduce the notation $\tilde{\partial}^{i}$ and $\tilde{x}_{i}$ not only for $i = a$ but also for $i = \mu$ by formally introducing $\tilde{x}_{\mu}$ and assuming that the fields do not depend on $\tilde{x}_\mu$, so practically $\tilde{\partial}^{\mu}=0$.

T-duality is now defined as a $GL(2d; \mathbb{Z})$ subclass of a diffeomorphism of the doubled torus $T^{d} \times T^{d}$ that preserves the $O(d,d)$ metric,
\begin{align}
 \eta_{MN}dX^{M} dX^{N} = 2 dx^{a} d\tilde{x}_{a}\,,
 \quad
 (X^M)=(x^a,\tilde{x}_a). 
\end{align}
T-dual transformations form $O(d,d; \mathbb{Z})$ group.
To construct an $O(d,d; \mathbb{Z})$ invariant action, we represent $O(d,d; \mathbb{Z}) \subset GL(2d; \mathbb{Z})$ transformations as a subgroup of $O(D,D; \mathbb{Z}) \subset GL(2D; \mathbb{Z})$ transformations.
For a given element of $O(D,D; \mathbb{Z})$, one can define a natural action for $h_{ij}$ and $b_{ij}$. We skip the details of this action but the important fact here is that the combination
$e_{i \bar{j}} := h_{ij } + b_{ij}$
is a covariant tensor of $O(D,D;\mathbb{Z})$, i.e.,
it transforms as $e_{i \bar{j}} \rightarrow M_{i}{}^{k} \bar{M}_{\bar{j}}{}^{\bar{l}} e_{k \bar{l}}$ with matrices $M_{i}{}^{k},  \bar{M}_{\bar{j}}{}^{\bar{l}}$ associated with the element of $O(D,D;\mathbb{Z})$ (see the review \cite{Zwiebach Review} for details).
The background value of $e_{i \bar{j}}$ is defined by $E^{bg}_{i \bar{j}} := g^{bg}_{ij}  + b^{bg}_{ij}$, where the background value of $b_{ij}$ is understood as $b^{bg}_{ij} = 0$ in our case.
We can also define the $O(D,D;\mathbb{Z})$ covariant derivative $D_{i}$ and $\bar{D}_{\bar{i}}$ by
\begin{equation}
D_{i} := \partial_{i} - (E^{bg})_{ij} \tilde{\partial}^j, \,\,\,\,\,\,\,\, \bar{D}_{\bar{i}} := \partial_{i} + (E^{bg})_{ji} \tilde{\partial}^{j},
\label{cov}
\end{equation}
which transform as $D_{i} \rightarrow M_{i}{}^{j} D_{j}$ and $\bar{D}_{\bar{i}} \rightarrow \bar{M}_{\bar{i}}{}^{\bar{j}} \bar{D}_{\bar{j}}$.
Another important fact is that the background metric $g^{bg}_{ij}$ transforms  as
$g^{bg}_{ij} \rightarrow M_{i}{}^{k} M_{j}{}^{l} g^{bg}_{kl} = \bar{M}_{\bar{i}}{}^{\bar{k}} \bar{M}_{\bar{j}}{}^{\bar{l}} g^{bg}_{\bar{k}\bar{l}}$. Thus the index structure of the background metric can be regarded as either $g^{bg}_{ij}$ or $g^{bg}_{\bar{i}\bar{j}}$.
Thus $O(d,d;\mathbb{Z})$ scalar can be obtained by contracting the covariant indices $i, j, \dots$ and $\bar{i}, \bar{j}, \dots$ by the inverse metric $g^{ij}_{bg}$ or $g^{\bar{i}\bar{j}}_{bg}$.
Note that $O(d,d;\mathbb{Z})$ scalars are automatically Lorentz scalars because $i,j, \dots$ and $ \bar{i}, \bar{j}, \dots$ are originally Lorentz indices.

To have a healthy massless theory, we also impose the following gauge symmetry:
\begin{align}
 \delta h_{ij} &= \partial_{i} \epsilon_{j} + \partial_{j} \epsilon_{i} + \tilde{\partial}_{i} \tilde{\epsilon}_{j} + \tilde{\partial}_{j} \tilde{\epsilon}_{i} \\
 \delta b_{ij} &= - (\partial_{i} \tilde{\epsilon}_{j} - \partial_{j} \tilde{\epsilon}_{i}) - (\tilde{\partial}_{i} \epsilon_{j} - \tilde{\partial}_{j} \epsilon_{i}) \\
 \delta d &= - \frac{1}{2} \partial_{i} \epsilon^{i} - \frac{1}{2} \tilde{\partial}_{i} \tilde{\epsilon}^{i},
\end{align}
where the gauge parameters $\epsilon_{i}$ and $\tilde{\epsilon}_{i}$ are functions of $\{x^{\mu}, x^{a}, \tilde{x}_{a}\}$.

By requiring the $O(d,d;\mathbb{Z})$ and Lorentz invariance, as well as the above gauge symmetry,
the quadratic action for $h_{ij}, b_{ij},$ and $d$ is
given by
\begin{align}
 S^{(2)}_{\text{DFT}} = \frac{1}{2 \kappa_{D}^2} \int [d x d \tilde{x}] {\cal L}^{(2)}_{DFT}
\end{align}
with
\begin{align}
 {\cal L}^{(2)}_{\text{DFT}} =&
\frac{1}{8} e^{i \bar{j}} (D^2 + \bar{D}^2) e_{i \bar{j}} + \frac{1}{4} ( \bar{D}^{\bar{j}} e_{i \bar{j}})^2 + \frac{1}{4} (D^{i} e_{i \bar{j}} )^2 -2d D^{i} \bar{D}^{ \bar{j} } e_{i \bar{j}} - 2 d (D^2 + \bar{D}^2) d \notag\\ 
=& 
\frac{1}{4} h^{ij} \partial^2 h_{ij} + \frac{1}{2} ( \partial^i h_{ij})^2 -2d \partial^{i} \partial^{j} h_{ij} -4 d \partial^2 d + \frac{1}{4} b^{ij} \partial^2 b_{ij} + \frac{1}{2} ( \partial^j b_{ij})^2 \notag\\
& + \frac{1}{4} h^{ij} \tilde{\partial}^2 h_{ij} + \frac{1}{2} ( \tilde{\partial}^i h_{ij})^2 + 2d \tilde{\partial}^{i} \tilde{\partial}^{j} h_{ij} -4 d \tilde{\partial}^2 d + \frac{1}{4} b^{ij} \tilde{\partial}^2 b_{ij} + \frac{1}{2} ( \tilde{\partial}^j b_{ij})^2  \notag\\
& +(\partial_{k} h^{ik}) ( \tilde{\partial}^j b_{ij}) + ( \tilde{\partial}^{k} h_{ik} ) ( \partial_{j} b^{ij}) -4d \partial^{i} \tilde{ \partial}^{j} b_{ij}.
\label{DFT}
\end{align}
Here $ [dx d\tilde{x}] := \left(  \prod_{\mu} dx^{\mu} \right)   \left(  \prod_{a}  dx^{a} d \tilde{x}_{a}  \right)$.
We note that the expression in Eq.~\eqref{DFT} has an ambiguity because 
\begin{equation}
D^2 - {\bar{D}}^2 = -4 \partial_{i} \tilde{\partial}^{i} = 0
\label{identity of cov}
\end{equation}
under the level-matching condition. Note that the gauge invariance cannot be maintained without imposing the level-matching condition. Also, if we decouple the winding modes, the above quadratic action is reduced to the following second order action:
\begin{align}
 S_{\text{DFT}}^{(2)}|_{\tilde{\partial} = 0} = \left. \frac{1}{2 \kappa_{D}^2} \frac{\left(2 \pi \alpha'\right)^{d}}{(R^a)^d} \int d^D x \sqrt{-g} \mathrm{e}^{- 2 \varphi} \left( R + 4 \partial_{i} \varphi \partial^{i} \varphi - \frac{1}{12} H_{ijk} H^{ijk} \right) \right|_{\text{second order}},
\end{align}
where $\varphi$ is the conventional dilaton defined by $\varphi := d + \frac{1}{4} h_{i}{}^{i}$ and $H_{ijk}$ is the field strength of $b_{ij}$: $H_{ijk} = 3 \partial_{[i} b_{jk]}$. 

\section{Massive deformation of DFT}
Now we study massive deformations of DFT by relaxing the assumption of gauge invariance. As we mentioned, gauge invariance is spoiled once we relax the level-matching condition. In particular, $D^2-{\bar{D}}^2$ does not vanish anymore when acting on the fields $e_{i\bar{j}}$ and $d$. Also, we can include mass terms in an $O(d,d; \mathbb{Z})$ invariant manner. There are two such mass terms: $e_{i\bar{j}} e^{i\bar{j}}$ and $d^2$.
Thus, we consider the following massive deformations\footnote{
At first glance, this Lagrangian might look similar to that in Ref.~\cite{Chen-Te Ma}, where a DFT of $(N_{R},N_{L})=(2,0)$ and $(0,2)$ excitations was constructed. However, they used the vector excitation $\alpha_{-2}^{i} \ket{0}$ to define the $b$-field and so the field contents are different 
from ours.}:
\begin{align}
\label{massive_deformations}
 S^{(2)}_{\text{mDFT}} &= \frac{1}{2 \kappa_{D}^2}\int [dx d\tilde{x}] {\cal L}^{(2)}_{\text{mDFT}}
\end{align}
with
\begin{align}
{\cal L}^{(2)}_{\text{mDFT}} &= {\cal L}^{(2)}_{\text{DFT}}  - \frac{1}{16} \zeta e^{i \bar{j}} (D^2 - \bar{D}^2) e_{i \bar{j}}  + \theta d (D^2 - \bar{D}^2) d - \frac{1}{4} m_e^2 e_{i \bar{j}} e^{i \bar{j}} + 4 m_{d}^2 d^2\,.
\label{mDFT}
\end{align}
Here the first term is the massless DFT Lagrangian~\eqref{DFT}. On top of it, we introduced four parameters $\zeta, \theta, m_e^2, m_{d}^2 \in \mathbb{R}$. At this moment, we do not impose any level-matching condition.
One may also write $\mathcal{L}^{(2)}_{\text{mDFT}}$ in terms of $h_{ij}$,$b_{ij}$ and $d$ as
\begin{align}
{\cal L}^{(2)}_{\text{mDFT}} =& 
{\cal L}_{\text{DFT}}^{(2)} + \frac{1}{4}\zeta (h^{ij} \partial_{k} \tilde{\partial}^{k} h_{ij} + b^{ij} \partial_k \tilde{\partial}^k b_{ij}) - 4 \theta  d \partial_{k} \tilde{\partial}^{k} d 
-  \frac{1}{4} m^2 (h^{ij} h_{ij} + b^{ij} b_{ij}) + 4 m_{d}^2 d^2 .
\label{expand mDFT}
\end{align}
In the following, we study the particle spectrum of the theory from the $n$-dimensional field theory point of view and clarify under which conditions the theory is free from ghosts and tachyons.

\subsection{Kaluza-Klein decomposition}
First, let us perform the following KK decomposition of $e_{i\bar{j}}$:
\begin{align}
(e_{i\bar{j}})=
\left(\begin{array}{cc}e_{\mu\nu} & \sqrt{2}A^+_{a\nu } \\ \sqrt{2}A^-_{b\mu }  &\sqrt{2}\Phi_{ab}\end{array}\right)\,,
\end{align}
where $e_{\mu\nu}=h_{\mu\nu}+b_{\mu\nu}$ in particular.
By substituting the Fourier decompositions (\ref{DoubleFourier}) to (\ref{massive_deformations}) and performing the integration along the compactified doubled space $T^{d} \times T^{d}$, we obtain the Fourier expansion of the action,
\begin{align}
 S_{\text{mDFT}}^{(2)} = \frac{( 2 \pi)^{2d} { \alpha' }^d}{2 \kappa_{D}^2} \int d^n x
\sum_{Z \in \mathbb{Z}^{2d}} {\cal L}_{\text{mDFT},Z}^{(2)}\,,
\end{align}
where $Z := (n_{a}, w^{a})$ are the KK level and the winding numbers. Also, ${\cal L}_{\text{mDFT},Z}^{(2)}$ is given by
\begin{align}
{\cal L}_{\text{mDFT},Z}^{(2)}
&=\frac{1}{4}
\left[
e^{\mu\nu*}\left(\Box-m^2\right)e_{\mu\nu}
\right]
+\frac{1}{2}\sum_\pm A^{\pm*}_{a\mu}\left(\Box-m^2\right)A^{\pm\mu}_{a}
+\frac{1}{2}\Phi^*_{ab}\left(\Box-m^2\right)\Phi^{ab}
\nonumber
\\
&\quad
-4d^*\left[\Box-m^2\right]d+4 \left[(\theta-\zeta)p\cdot\tilde{p}+(m_d^2-m_e^2)\right]|d|^2
\nonumber
\\
&\quad
+\frac{1}{4}\left|\partial^\mu e_{\mu\nu}+\sqrt{2}ip_-^aA^-_{a\nu}\right|^2
+\frac{1}{4}\left|\partial^\nu e_{\mu\nu}+\sqrt{2}ip_+^bA^+_{b\mu}\right|^2
\nonumber
\\
&\quad
+\frac{1}{2}\left|\partial^\mu A^+_{b\mu}+ip_-^a\Phi_{ab}\right|^2
+\frac{1}{2}\left|\partial^\nu A^-_{a\nu}+ip_+^b\Phi_{ab}\right|^2
\nonumber
\\
\label{KK}
&\quad
-d^*\left[
\partial^\mu\partial^\nu e_{\mu\nu}
+\sqrt{2}ip_+^b\partial^\mu A^+_{b\mu}
+\sqrt{2}ip_-^a\partial^\nu A^-_{a\nu}
-\sqrt{2}p_-^ap_+^b\Phi_{ab}
\right]+{\rm c.c.}
\,,
\end{align}
where we defined $p_\pm^a$ and $m^2(p,\tilde{p})$ by
\begin{align}
p_\pm^a := p^a \pm \tilde{p}^a\,,
\quad
m^2(p,\tilde{p}):=m_e^2+p^2+\tilde{p}^2+\zeta p\cdot\tilde{p}\, .
\end{align}
Here we use the notation $p \cdot \tilde{p} := p^{a} \tilde{p}_{a},\, p^2:= p^{a} p_{a}$ and so on.
Later, $m^2(p,\tilde{p})$ will be identified with the mass squared of the mode with the KK momenta $p^a$ and the winding numbers $\tilde{p}^a$ (for notational simplicity, we often suppress the $p,\tilde{p}$-dependence).
Also note that in Eq.~\eqref{KK} nontrivial mixings appear only in the last three lines. To resolve these mixings and diagonalize the Lagrangian, it is convenient to perform the following tensor decomposition:
\begin{align}
e^T_{\mu\nu}&:=\left( P^{\perp}_{\mu}{}^{\rho} P^{\perp}_{\nu}{}^{\sigma} - \frac{1}{n-1} P^{\perp}_{\mu\nu} P^{\perp \rho\sigma} \right) e_{\rho\sigma}\,,
\quad
e^{-V}_{\mu\nu}:=P^{\perp}_{\mu}{}^{\rho}P^{\parallel}_{\nu}{}^{\sigma}e_{\rho\sigma}
\,,
\quad
e^{+V}_{\mu\nu}:=P^{\parallel}_{\mu}{}^{\rho}P^{\perp}_{\nu}{}^{\sigma}e_{\rho\sigma}
\,,
\nonumber
\\
h_\parallel&:=P^{\parallel\mu\nu}e_{\mu\nu}
\,,
\quad
h_\perp:=P^{\perp\mu\nu}e_{\mu\nu}\,,
\quad
A^{\pm V}_{a\mu}:=P^{\perp}_{\mu}{}^{\nu}A^\pm_{a\nu}\,,
\quad
A^{\pm S}_{a\mu}:=P^{\parallel}_{\mu}{}^{\nu}A^\pm_{a\nu}\,,
\end{align}
where we defined the transverse and longitudinal projectors by
\begin{align}
 P^{\perp}_{\mu}{}^{\rho} := \delta_{\mu}{}^{\rho}-\frac{ \partial_{\mu}\partial^{\rho}}{\Box}\,,
 \quad
 P^{\|}_{\mu}{}^{\rho} := \frac{\partial_{\mu}\partial^{\rho}}{\Box}\,.
\end{align}
In the rest of the section, we compute the spectrum of each sector.

\subsection{Tensor sector}

The Lagrangian of the tensor sector reads
\begin{align}
 {\cal L}^{T}_{Z} = \frac{1}{4}e_{\mu\nu}^{T*}\left(\Box-m^2\right)e^{T\mu\nu}\,.
\end{align}
In terms of  the tensor components of the metric $h^T_{\mu\nu}$ and $b$-field $b^T_{\mu\nu}$, we may rewrite it as
\begin{align}
 {\cal L}^{T}_{Z} = \frac{1}{4} h^{T *}_{\mu\nu} \left(\Box-m^2\right) h^{T\mu\nu} 
 + \frac{1}{4} b^{T*}_{\mu\nu} \left(\Box-m^2\right)b^{T \mu\nu} \,,
\end{align}
which describes a massive spin $2$ particle and a massive anti-symmetric tensor with the mass $m$. To avoid tachyonic instability, we require
\begin{align}
m^2(p,\tilde{p})>0\,.
\end{align}

\subsection{Vector sector}

To discuss the vector sector, it is convenient to decompose the vectors $A^{\pm V}_{a\mu}$ with respective to the internal cycle indices as\footnote{
We focus on the sector $p_\pm^2\neq0$, since the analysis for $p_\pm^2=0$ is trivial.}
\begin{align}
\label{A_a_decompose}
A_{a\mu}^{\pm V}=\widehat{A}_{a\mu}^{\pm V}+\frac{p_{\pm a}}{p_\pm}\mathcal{A}_{\mu}^{\pm V}
\quad
{\rm with}
\quad
\widehat{A}_{a\mu}^{\pm V}:=\left(\delta_{a}^b-\frac{p_{\pm a}p_\pm^b }{p_\pm^2}\right)A_{b\mu}^{\pm V}\,,
\quad
\mathcal{A}_{\mu}^{\pm V}:=\frac{p_\pm^a}{p_\pm}A_{a\mu}^{\pm V}\, ,
\end{align}
where $p_{\pm}$ without indices are defined by
\begin{align}
 p_{\pm} := \sqrt{p_{\pm}^{a} p_{\pm a}} = \sqrt{p^2 + \tilde{p}^2 \pm 2 p \cdot \tilde{p}} .
\end{align}
Noticing that only $\mathcal{A}_\mu^{\pm V}$ mix with non-dynamical fields $e^{\pm V}_{\mu\nu}$, we find
\begin{align}
{\cal L}^{V}_{Z} &=
\frac{1}{2}\sum_{\pm}
\left[
\widehat{A}^{\pm V*}_{a\mu}\left(\Box-m^2\right)\widehat{A}^{\pm V\mu}_{a}
+\frac{m^2-p_\pm^2}{m^2}\mathcal{A}^{\pm V*}_{\mu}\left(\Box-m^2\right)\mathcal{A}^{\pm V\mu}
\right]
\nonumber
\\
&\quad
-\frac{m^2}{4}\left|e^{+ V}_{\mu\nu}+\sqrt2i\frac{p_-}{m^2}\partial_\mu \mathcal{A}^{-V}_{\nu}\right|^2
-\frac{m^2}{4}\left|e^{- V}_{\mu\nu}+\sqrt2i\frac{p_+}{m^2}\partial_\nu \mathcal{A}^{+V}_{\mu}\right|^2
\,.
\end{align}
Integrating out $e^{\pm V}_{\mu\nu}$ gives
\begin{align}
{\cal L}^{V}_{Z} &=
\frac{1}{2}\sum_{\pm}
\left[
\widehat{A}^{\pm V*}_{a\mu}\left(\Box-m^2\right)\widehat{A}^{\pm V\mu}_{a}
+\frac{m^2-p_\pm^2}{m^2}\mathcal{A}^{\pm V*}_{\mu}\left(\Box-m^2\right)\mathcal{A}^{\pm V\mu}
\right]\,,
\end{align}
which contains $2d$ massive spin $1$ particles with the mass $m$. Therefore, the vector sector is free from ghosts and tachyons if and only if
\begin{align}
\label{no-ghost_no-tachyons}
m^2(p,\tilde{p})-p_+^2>0\,,
\quad
m^2(p,\tilde{p})-p_-^2>0\,.
\end{align}
Note that the tachyon free condition $m^2(p,\tilde{p})>0\, $ is automatically satisfied under Eq.~\eqref{no-ghost_no-tachyons}.

\subsection{Scalar sector}
Similarly to Eq.~\eqref{A_a_decompose}, let us decompose the scalar components with internal cycle indices as
\begin{align}
&\widehat{A}_{a\mu}^{\pm S}:=\left(\delta_{a}^b-\frac{p_{\pm a}p_\pm^b }{p_\pm^2}\right)A_{b\mu}^{\pm S}\,,
\quad
\mathcal{A}_{\mu}^{\pm S}:=\frac{p_\pm^a}{p_\pm}A_{a\mu}^{\pm S}\,,
\quad
\widehat{\Phi}_{ab}:= \left(\delta_{a}^c-\frac{p_{- a}p_-^c }{p_-^2}\right)\left(\delta_{b}^d-\frac{p_{+ b}p_+^d }{p_+^2}\right)\Phi_{cd}\,,
\nonumber
\\
&
\widehat{\Phi}_{-b}:=\frac{p_-^a}{p_-}\left(\delta_{b}^d-\frac{p_{+ b}p_+^d }{p_+^2}\right)\Phi_{ad}\,,
\quad
\widehat{\Phi}_{a+}:=\left(\delta_{a}^c-\frac{p_{- a}p_-^c }{p_-^2}\right)\frac{p_+^b}{p_+}\Phi_{cb}\,,
\quad
\Phi_{-+}:=\frac{p_-^ap_+^b}{p_- p_+}\Phi_{ab}\,.
\end{align}
Together with $h_\parallel$, $h_\perp$, and $d$, we have now 9 types of scalar components. To avoid complication, we further classify them into the following two subsectors that are decoupled from each other:
\begin{enumerate}
\item $h_\perp$, $\widehat{A}_{a\mu}^{\pm S}$, $\widehat{\Phi}_{ab}$, $\widehat{\Phi}_{-b}$, $\widehat{\Phi}_{a+}$,
\item $h_\parallel$, $d$, $\mathcal{A}_\mu^{\pm S}$, $\Phi_{-+}$.
\end{enumerate}

\paragraph{Subsector 1.}

The Lagrangian ${\cal L}^{S_1}_{Z}$ of the first subsector reads
\begin{align}
{\cal L}^{S_1}_{Z} &=
\frac{1}{4(n-1)}h_\perp^*\left(\Box-m^2\right)h_\perp
+\frac{1}{2}\widehat{\Phi}_{ab}^*\left(\Box-m^2\right)\widehat{\Phi}_{ab}
\nonumber
\\
&\quad
+\frac{1}{2}\frac{m^2-p_-^2}{m^2}\widehat{\Phi}_{-b}^*\left(\Box-m^2\right)\widehat{\Phi}_{-b}-\frac{m^2}{2}\left|\widehat{A}_{b\mu}^{+S}+i\frac{p_-}{m^2}\partial_\mu\widehat{\Phi}_{-b}\right|^2\,,
\nonumber
\\
&\quad
+\frac{1}{2}\frac{m^2-p_+^2}{m^2}\widehat{\Phi}_{a+}^*\left(\Box-m^2\right)\widehat{\Phi}_{a+}
-\frac{m^2}{2}\left|\widehat{A}_{a\nu}^{-S}+i\frac{p_+}{m^2}\partial_\nu\widehat{\Phi}_{a+}\right|^2\,.
\end{align}
Integrating out non-dynamical fields $\widehat{A}_{a\mu}^{\pm S}$, we arrive at
\begin{align}
{\cal L}^{S_1}_{Z} &=
\frac{1}{4(n-1)}h_\perp^*\left(\Box-m^2\right)h_\perp
+\frac{1}{2}\widehat{\Phi}_{ab}^*\left(\Box-m^2\right)\widehat{\Phi}_{ab}
\nonumber
\\
&\quad
+\frac{1}{2}\frac{m^2-p_+^2}{m^2}\widehat{\Phi}_{a+}^*\left(\Box-m^2\right)\widehat{\Phi}_{a+}
+\frac{1}{2}\frac{m^2-p_-^2}{m^2}\widehat{\Phi}_{-b}^*\left(\Box-m^2\right)\widehat{\Phi}_{-b}
\,,
\end{align}
which describes $d^2$ massive scalars with the mass $m$. We find that absence of ghosts and tachyons in this sector requires the same conditions as Eq.~\eqref{no-ghost_no-tachyons}.

\paragraph{Subsector 2.}

To discuss the other sector, it is convenient to define
\begin{align}
\phi :=\frac{1}{\sqrt2}(h_\parallel+4d)\,.
\end{align}
In this language, the Lagrangian reads
\begin{align}
{\cal L}^{S_2}_{Z} &=
-\frac{1}{2}\phi^*\left(\Box+m^2\right)\phi
+\frac{1}{2}\Phi_{-+}^*\left(\Box-m^2\right)\Phi_{-+}
+\frac{1}{2}\left(p_-^2+p_+^2\right)|\Phi_{-+}|^2
\nonumber
\\
&\quad
+4 \left[(\theta-\zeta)p\cdot\tilde{p}+(m_d^2-m_e^2)\right]|d|^2
+\sqrt{2}
\left[
d^*\Big(m^2\phi + p_+ p_-\Phi_{-+}\Big)+{\rm c.c.}
\right]
\nonumber
\\
&\quad
+\frac{1}{2}\frac{1}{m^2-p_+^2}\Big| p_+\partial_\mu\phi+ p_- \partial_\mu\Phi_{-+}\Big|^2
+\frac{1}{2}\frac{1}{m^2-p_-^2}\Big| p_- \partial_\mu\phi+ p_+ \partial_\mu\Phi_{-+}\Big|^2
\nonumber
\\
&\quad
-\frac{m^2-p_+^2}{2}\Big|\mathcal{A}_\mu^{+S}+\frac{i}{m^2-p_+^2}\Big( p_+ \partial_\mu\phi+ p_- \partial_\mu\Phi_{-+}\Big)\Big|^2
\nonumber
\\
&\quad
-\frac{m^2-p_-^2}{2}\Big|\mathcal{A}_\mu^{-S}+\frac{i}{m^2-p_-^2}\Big( p_-\partial_\mu\phi+ p_+ \partial_\mu\Phi_{-+}\Big)\Big|^2\,.
\end{align}
First, integrating out non-dynamical fields $\mathcal{A}_\mu^{\pm S}$ gives
\begin{align}
{\cal L}^{S_2}_{Z} &=
-\frac{1}{2}\phi^*\left(\Box+m^2\right)\phi
+\frac{1}{2}\Phi_{-+}^*\left(\Box-m^2\right)\Phi_{-+}
+\frac{1}{2}\left(p_-^2+p_+^2\right)|\Phi_{-+}|^2
\nonumber
\\
&\quad
+4 \left[(\theta-\zeta)p\cdot\tilde{p}+(m_d^2-m_e^2)\right]|d|^2
+\sqrt{2}
\left[
d^*\Big(m^2\phi + p_+ p_- \Phi_{-+}\Big)+{\rm c.c.}
\right]
\nonumber
\\
&\quad
+\frac{1}{2}\frac{1}{m^2-p_+^2}\Big| p_+ \partial_\mu\phi+ p_- \partial_\mu\Phi_{-+}\Big|^2
+\frac{1}{2}\frac{1}{m^2-p_-^2}\Big| p_- \partial_\mu\phi+ p_+ \partial_\mu\Phi_{-+}\Big|^2
\,,
\end{align}
where notice that the kinetic term of $\phi$ in the first line has a wrong sign. As it suggests, one can explicitly show that there appears a ghost for generic values of the model parameters. The only way to remove the ghost is to tune the parameters such that
\begin{align}
\label{FP-like}
(\theta-\zeta)p\cdot\tilde{p}+(m_d^2-m_e^2)=0\,,
\end{align}
which is analogous to the ghost-free condition in the Fierz-Pauli theory. Under this condition, the equation of motion for $d$ gives a constraint,
\begin{align}
m^2\phi  + p_+ p_-\Phi_{-+}=0\,,
\end{align}
and so the Lagrangian after integrating out the dilaton $d$ reads
\begin{align}
{\cal L}^{S_2}_{Z} &=
\frac{1}{2}\frac{m^2-p_-^2}{m^2}\frac{m^2-p_+^2}{m^2}\Phi_{-+}^*\left(\Box-m^2\right)\Phi_{-+}
\,,
\end{align}
which describes a massive scalar with a correct sign of the kinetic term if we assume~\eqref{no-ghost_no-tachyons}.

\bigskip
To summarize, the scalar sector is free from ghosts and tachyons if and only if both of the conditions~\eqref{no-ghost_no-tachyons} and~\eqref{FP-like} are satisfied. Under these conditions, there exist $d^2+1$ massive scalars with the identical mass $m$.

\subsection{Implications}
To summarize all the results above, our massive DFT is free from ghosts and tachyons if and only if the conditions~\eqref{no-ghost_no-tachyons} and~\eqref{FP-like} are satisfied.

Under the condition~\eqref{FP-like}, we find that all the particles have the same physical mass $m^2(p ,\tilde{p}) = m_e^2+p^2+\tilde{p}^2+\zeta p\cdot\tilde{p}$. This is consistent with the results obtained by Olaf et al.~\cite{Curious}, where the level matching condition $p\cdot \tilde{p} = 0$ is imposed.  
Their analysis is at the level of the equations of motion, while we derived the action for each helicity mode. Hence, we can
see the absence of ghost explicitly.
By plugging the expressions of $m(p,\tilde{p})$ and $p_{\pm}$, the conditions \eqref{no-ghost_no-tachyons} read
\begin{eqnarray}
&&  m_{e}^2 + (\zeta-2) p \cdot \tilde{p} >0, \qquad     m_{e}^2 + (\zeta+2) p \cdot \tilde{p} >0.
\label{healthy conditions}
\end{eqnarray}
It is easy to see that the level matching condition $p \cdot \tilde{p} = 0$, with a positive $m^2_{e}$,  is a sufficient condition for the absence of ghosts as well as tachyons. Interestingly, there are other healthy theories with $p \cdot \tilde{p} \neq 0$. 
In Figure 1, we depicted the stable parameter region in the $( \zeta, \frac{p \cdot \tilde{p}}{m^2_{e}} )$ plane where
ghosts and tachyons are absent. 
 \begin{figure}[h]
	\centering
	\includegraphics[width=10cm]{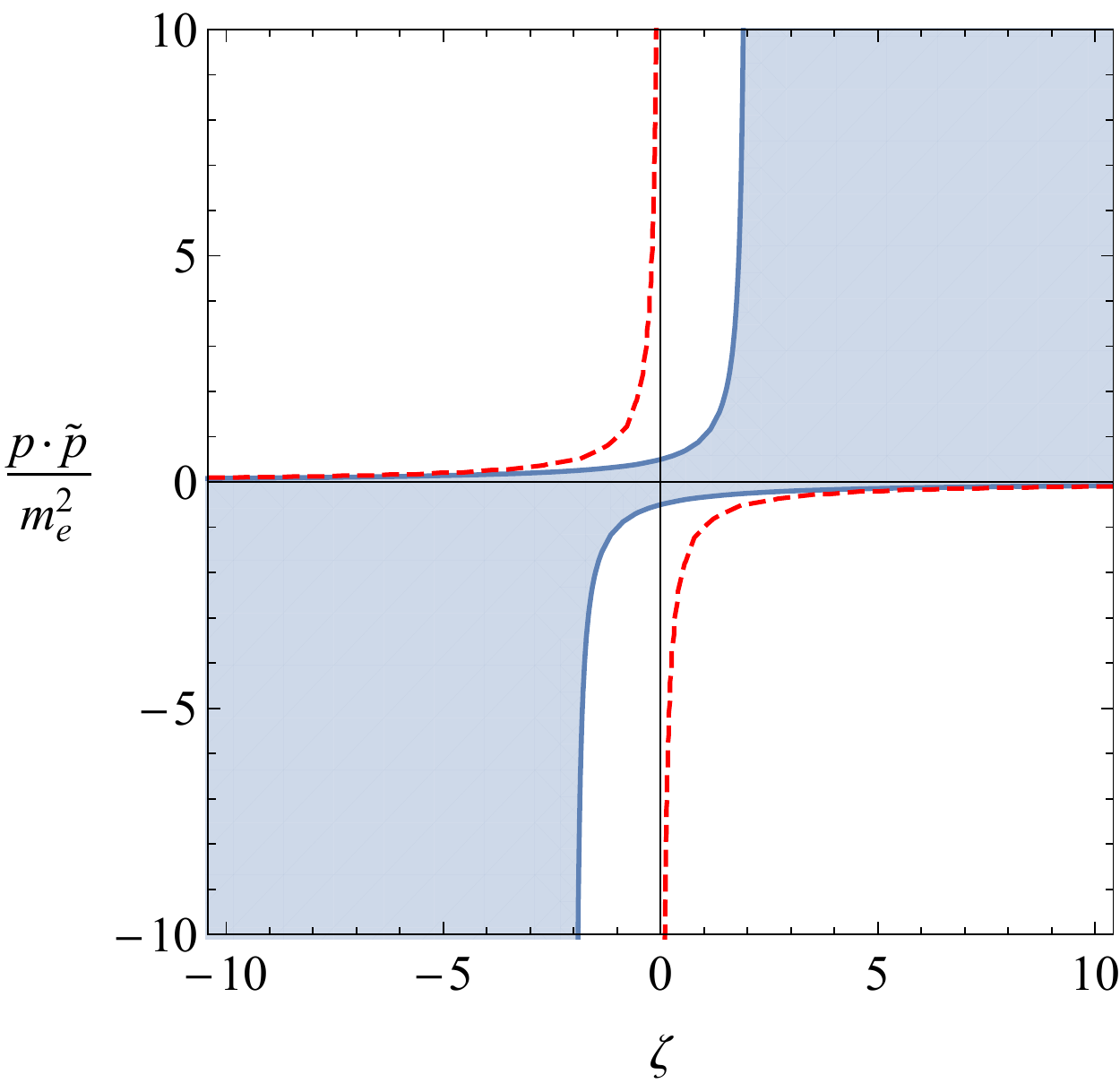}
	\caption{Ghost and tachyon free conditions in massive DFT. The shaded region represents
	ghost and tachyon free parameters. The dashed lines represent the massless modes in the entire $D$ dimensional target space.}
	\label{Fig1}
\end{figure}
When we build a theory consisting of a symmetric rank 2 tensor, a 2 form field and a scalar, assuming $O(d,d;\mathbb{Z})$  symmetry only  (we abandon diffeomorphism by introducing the bare mass and relaxing the constraint $p \cdot \tilde{p} =0$), Figure 1 tells us what values of $\zeta$ and $\frac{p \cdot \tilde{p}}{m^2_{e}}$ are allowed. 
For example, for a given parameter $\zeta > 2$, the almost all of the negative $p \cdot \tilde{p}$ are excluded. In this sense, massive DFT with $\zeta > 2$ requires a kind of ``level-matching condition'' that limits the value of $p \cdot \tilde{p}$ positive.
Similar discussion holds for other choices of parameter $\zeta$.

Since $m^2$ includes a possibly negative contribution $\zeta p \cdot \tilde{p}$, apparently it looks possible to obtain massless spectrum where the $ \zeta p \cdot \tilde{p}$ term cancels the effect of bare mass term.
However, by defining the target space mass by $\widehat{m}^2 := m^2 - (p^2 + \tilde{p}^2)$, we can derive 
a lower bound on $\widehat{m}^2$ as
\begin{align}
 \widehat{m}^2 = m^2_{e} + \zeta p \cdot \tilde{p} > 2 |p \cdot \tilde{p}|.\label{targetmass}
\end{align}
One can immediately find that the vanishing target mass is not allowed in massive DFT.
Actually, the parameter for the vanishing target mass is represented as the dashed line in Figure~\ref{Fig1}, which is out of the stable parameter region.

As another implication of Eq.~\eqref{targetmass}, we mention the stringy UV completion of massive gravity. 
If we assume that massive gravity can be embedded into string theory, the right hand side of Eq.~\eqref{targetmass} should be of the order of the string scale because
\begin{align}
 2 | p \cdot \tilde{p} | = \frac{2}{\alpha'} |n_{a} w^{a}|, \qquad n_a w^{a} \in \mathbb{Z}.
\end{align}
Hence, the only way to get a smaller mass than the string scale is to impose the weak constraint $ p \cdot \tilde{p} =0 $ for all modes. 
Otherwise, we are forced to consider the mass of order of the string scale. This point is also discussed in \cite{Curious}.

\section{Conclusion}
In this paper we studied massive deformations of DFT at the free theory level. Our starting point was the Lagrangian~\eqref{massive_deformations} with four parameters $\zeta, \theta, m_{e}^2, m_{d}^2$ without imposing any level-matching condition.
We find that the theory is free from ghosts and tachyons if and only if the conditions~\eqref{no-ghost_no-tachyons} and~\eqref{FP-like} are satisfied. The condition \eqref{FP-like} reduces the four parameters of theory to two parameters, $\zeta$ and $m_{e}^2$.
The conditions~\eqref{no-ghost_no-tachyons}, which can be written explicitly as~\eqref{healthy conditions}, are understood as conditions analogous to the weak constraint: For a given parameter $\zeta$, the consistency conditions~\eqref{healthy conditions} give a bound on $p \cdot \tilde{p}$. Besides, we demonstrated that the standard weak constraint $p \cdot \tilde{p} =0$ is picked up if we require that the mass of the lightest massive spin 2 particle is lighter than the string scale, which is relevant when exploring stringy UV completion of massive gravity in the regime of phenomenological interests.

Among others, the most important future direction is to generalize our analysis to interacting theories. As we mentioned in introduction, the present formulation of DFT relies on the strong constraint, which ensures gauge invariance, but the winding modes are projected out at this cost: without relaxing the strong constraint, one cannot discuss phenomenology of winding modes. Since massive theories are realized in the gauge symmetry broken phase, construction of a consistent massive DFT could be a bypass to this issue. The nontriviality there is in identifying the ghost-free conditions at the interacting level. A next step in this direction will be to embed dRGT massive gravity~\cite{deRham:2010ik, Resummation of MG} into the DFT framework and clarify if the strong constraint is required for the theory to be ghost-free. We hope to report our progress in this direction in the near future.

\section*{Acknowledgements}

T.~N.\ and J.~S.\  are supported in part by JSPS KAKENHI Grant Numbers JP17H02894 and JP20H01902.
 D.~Y. is supported by JSPS Postdoctoral Fellowships No. 201900294 and JSPS KAKENHI Grant Numbers 19J00294 and 20K14469.

\end{document}